\documentclass[aps,twocolumn]{revtex4}
\usepackage{amsmath}
\usepackage{graphicx}

\begin{document}

\title{Global Persistence Exponent in Critical Dynamics: Finite Size induced Crossover.}

\author{
D.Chakraborty \footnote{email: tpdc2@mahendra.iacs.res.in} and
J.K.Bhattacharjee \footnote{email: tpjkb@mahendra.iacs.res.in}}
\affiliation{ Department of Theoretical Physics,\\
Indian Association for the Cultivation of Science,\\
Jadavpur, Kolkata-700032, India.}

\begin{abstract}
We extend the definition of a global order parameter to the case of a critical system confined between two infinite parallel plates separated by a finite distance $L$. For a quench to the critical point we study the persistence property of the global order parameter and show that there is a crossover behaviour characterized by a non universal exponent which depends on the ratio of the system size to a dynamic length scale.

\end{abstract}

\maketitle

Global persistence exponent for non-equilibrium critical dynamics
was introduced a decade ago\cite{1}, following the emergence of
similar exponents in the evolution of Ising spins \cite{2}-\cite{4}
in one and higher dimensions and the evolution of a diffusing field
\cite{5} from random initial conditions in different dimensions. The
simplest system exhibiting persistence is the random walk in one
dimension \cite{6}. Since Brownian motion under restrictive geometry
has been of experimental interest lately \cite{7}, the persistence
problem was addressed under those situations \cite{8}. It was seen
that the power law decay for the infinite system acquired an
exponential correction for the confined system (confinement by walls
or harmonic forces). This was in contrast to the finite persistence
probability observed by Manoj and Ray for finite size systems
exhibiting critical dynamics. The quench carried out by Manoj and
Ray \cite{9} was, however, deep into the ordered region. For a
D-dimensional Ising model, starting from a random initial condition,
they quenched the system to $T=0$ and allowed the spins to evolve
according to Glauber dynamics. Domains began forming and when it
happened that the domain size became larger than the system size ,
then the persistence probability attained a finite value. The global
persistence exponent of Majumdar et al \cite{1} was defined
differently. It referred to the quench from a high temperature to
$T=T_c$, the critical point of the system and considered the global
order parameter. The individual spins flip rapidly and the
probability of not flipping in an interval has an exponential tail.
It is only when the global order parameter is considered, that one
finds the power law tail. In this situation if we consider a finite
size system, then for a sufficiently small system size (smaller than
the appropriate "dynamical" length scale), the global order
parameter will no longer find it so difficult to "overturn" and an
exponential tail could be expected just as happened with the
Brownian motion in restrictive geometry. In this note we use the
spherical limit to establish our result.\\
 We consider the usual
Landau Ginzburg free energy $F$ for the $N$-component order
parameter $\phi_{i}$ $\{i=1,2,....N\}$, in a 3-dimensional space,
that is ,
\begin{equation}\label{1}
F=\int \mathrm{d}^3 \vec{x}[{r \over 2} \phi_i \phi_i +{1 \over 2}
(\nabla_j \phi_i)(\nabla_j \phi_i) + {u \over N} (\phi_i
\phi_i)(\phi_j \phi_j)].
\end{equation}

The corresponding Langevin equation is given by
\begin{equation}\label{2}
 \dot{\phi_i}= \Gamma \nabla^2 \phi_i -\Gamma (r+{u \over N} \phi_i \sum_j
 \phi_j^{2})+\xi_i,
\end{equation}
where $\xi$ is a Gaussian white having correlation
\begin{equation}\label{3}
 <\xi(\vec{r},t) \xi(\vec{r'},t')> = 2\Gamma \delta(\vec{r}-\vec{r'})\delta(t-t').
\end{equation}
Since we will be using spherical limit, it makes sense to work in
$D=3$ directly. The range of validity of the spherical approximation
if for $2<D<4$ and hence $D=3$ is the natural choice. \\
The confinement is taken to be in the $z$-direction and the
orthogonal space has two dimensions. The confining is in the form of
two "parallel plates" at $z=0$ and at $z=L$, where Dirichlet
boundary conditions hold \cite{10}-\cite{12} . The other two dimensions are infinitely
extended. The decomposition of $\phi_i(\vec{r},t)$ is now in terms
of Fourier transform in two dimensions and a Fourier series in the
$z$-direction, so that
\begin{equation}\label{4}
 \phi_i(\vec{r},t) = \int {\mathrm{d}^2 \vec{k} \over (2\pi)^2} \sum_{n=1}^{\infty} \phi_{i,n} (\vec{k},t)
 \sin({n\pi z \over L}),
\end{equation}
and the linearized Langevin equation becomes
\begin{equation}\label{5}
\dot{\phi}_{i,n}(\vec{k})=-\Gamma (k^2+{n^2 \pi^2 \over L^2})
\phi_{i,n}(\vec{k})-\Gamma r \phi_{i,n}(\vec{k}),
\end{equation}
in the non-interacting limit,$u=0$. For the choice of $n=1$,
$r=-\frac{\pi^2}{L^2}$, $\vec{k}=0$ gives us
\begin{equation}\label{6}
\dot{\phi}_{i,1}(0)=\xi_i.
\end{equation}
At the critical point for the confined system ($r=-\pi^2/L^2$
represents the mean field expression of the critical point), the
lowest mode ($k=0$,$n=1$) undergoes a Brownian motion, corresponding
to a persistence exponent $\theta=0.5$. For the finite size system,
we identify $\phi_{i,1}(0)$ as the global order parameter.

To work in the spherical limit we write Eq.(\ref{2}) as
\begin{eqnarray}\label{7}
\nonumber
\dot{\phi}_{i,n}(\vec{k})= -\Gamma (k^2+\frac{n^2
\pi^2}{L^2})
\phi_{i,n}(\vec{k}) -\Gamma(r+\frac{u}{N}N <\phi^2>) \\
\phi_{i,n}(\vec{k}) +u \mathcal{O}(1/N)
\phi_{i,n}(\vec{k})+\xi_i(\vec{k},n,t).
\end{eqnarray}
Since $(N<\phi^2>-\sum_j \phi_j^2)$ is of $\mathcal{O}(1)$ and hence
in the limit $N \rightarrow \infty$ (spherical limit), we have, for
any $i$,
\begin{equation}\label{8}
\dot{\phi}_n(\vec{k})=-\Gamma \biggr( k^2+\frac{n^2
\pi^2}{L^2}\biggr) \phi_n(\vec{k})-\Gamma (r+u<\phi^2>)
\phi_n(\vec{k})+\xi(\vec{k},t).
\end{equation}
Defining $a(t)=-\Gamma(r+u<\phi^2>)$, we can write
\begin{equation}\label{9}
\dot{\phi}_n(\vec{k})=-\Gamma (k^2+\frac{n^2 \pi^2}{L^2})
\phi_n(\vec{k})+a(t) \phi_n(\vec{k})+\xi(\vec{k},t).
\end{equation}
The solution for $\phi_n(\vec{k},t)$ can now be written as
\begin{eqnarray}\label{10}
\nonumber 
\phi_n(\vec{k},t)&=&e^{-\Gamma(k^2+\frac{n^2
\pi^2}{L^2})t+b(t)}\biggr[\int_0^t \mathrm{d}t'e^{\Gamma
(k^2+\frac{n^2 \pi^2}{L^2})t'-b(t')}\\
& &\xi(\vec{k},t')\biggr] +\phi_n(\vec{k},0)
e^{-\Gamma(k^2+\frac{n^2 \pi^2}{L^2})t+b(t)},
\end{eqnarray}
where $b(t)=\int_0^t dt' a(t')$.
The long time dynamics is dominated
by the noise containing term and $<\phi^2>$ in that limit is given
by,
\begin{equation}\label{11}
<\phi^2>=\frac{2\Gamma}{g(t)} \sum_{k,n} \int_0^t \mathrm{d}t'
e^{-2\Gamma(k^2+\frac{n^2 \pi^2}{L^2})(t-t')}g(t'),
\end{equation}
where
\begin{equation}\label{12}
g(t)=e^{-2b(t)}.
\end{equation}
The dynamics of $g(t)$ is given by
\begin{eqnarray}\label{13}
\nonumber
  \dot{g} &=& -2g\dot{b}(t)=-2ga(t), \\
\nonumber
          &=&2g\Gamma(r+u<\phi^2>),  \\
\nonumber
          &=& 2r\Gamma g + 4u\Gamma  \int_0^t \mathrm{d}t'g(t')
\sum_{k,n}e^{-2\Gamma(k^2+\frac{n^2 \pi^2}{L^2})(t-t')}.\\
\end{eqnarray}
The critical point is now defined by the zero of the coefficient of
the $k=0$,$n=1$ component of $\phi_n(\vec{k})$ in Eq.(\ref{8}) and
thus
\begin{equation}\label{14}
 r_c+u<\phi^2>=-\frac{\pi^2}{L^2}.
\end{equation}
If we consider the Eq.(\ref{13}) at the critical point, then in the
terms of the Laplace transform
\begin{equation*}
 \tilde{g}(s)=\int_0^{\infty} g(t) e^{-st} \mathrm{d}t,
\end{equation*}
with
\begin{equation*}
 \int_0^{\infty} \dot{g}(t) e^{-st} \mathrm{d}t = s\tilde{g}(s)-1,
\end{equation*}
we arrive at
\begin{equation}\label{15}
\tilde{g}(s)=1/[s+\frac{2\Gamma \pi^2}{L^2}+4\Gamma^2
u\{\tilde{J}(0,L)-\tilde{J}(s,L)\}],
\end{equation}
where
\begin{equation}\label{16}
\tilde{J}(s,L)=\sum_{k,n} \frac{1}{s+2\Gamma(k^2+\frac{n^2
\pi^2}{L^2})}.
\end{equation}

\begin{eqnarray}\label{17}
\nonumber
  \triangle \tilde{J} &=& \tilde{J}(0,L)-\tilde{J}(s,L),\\
\nonumber
  &=&\sum_{k,n\geq 1}\frac{s}{s+2\Gamma(k^2+\frac{n^2
  \pi^2}{L^2})}
  \frac{1}{2\Gamma(k^2+\frac{n^2\pi^2}{L^2})}, \\
  \nonumber
  &=& \sum_{k,n\geq 0}\frac{s}{s+2\Gamma(k^2+\frac{n^2
  \pi^2}{L^2})}\frac{1}{2\Gamma(k^2+\frac{n^2\pi^2}{L^2})}\\
  &&-\int \frac{\mathrm{d}^2 \vec{k}}{(2\pi)^2} \frac{s}{(s+2\Gamma k^2)2\Gamma
  k^2}.
\end{eqnarray}
For $L\rightarrow \infty$ we can write
\begin{eqnarray}\label{18}
 \nonumber
  \triangle \tilde{J} &=& \frac{L}{\pi} \int \frac{\mathrm{d}^2 \vec{k}}{(2\pi)^2}
  \mathrm{d}\tilde{k}\frac{s}{s+2\Gamma(k^2+\tilde{k}^2)}
  \frac{1}{2\Gamma(k^2+\tilde{k}^2)} \\
\nonumber
   &=& \frac{L}{(2\pi)}\frac{4\pi}{(2\pi)^2}
   \int \frac{s k^2 \mathrm{d}^2 k}{(s+2\Gamma k^2)2\Gamma
   k^2}=\frac{L}{8\pi \Gamma}\bigr(\frac{s}{2\Gamma}\bigr)^{1/2}\\
   \end{eqnarray}

The first correction to $\triangle \tilde{J}$ for $L\rightarrow
\infty$ is given by the second term in Eq.(\ref{17}) which becomes
\begin{eqnarray}\label{19}
\nonumber
\int \frac{\mathrm{d}^2
\vec{k}}{(2\pi)^2}\frac{s}{(s+2\Gamma k^2)(2\Gamma k^2)} &=&
  \frac{s}{2\pi} \int \frac{k \mathrm{d}k}{(s+2\Gamma k^2)(2\Gamma k^2)}  \\
  \nonumber
  &=& \frac{s}{8\pi \Gamma} \int \frac{\mathrm{d}z}{z(z+\frac{s}{2\Gamma})} \\
  \nonumber
  &=& \frac{1}{8\pi \Gamma} \ln \biggr( \frac{z}{z+s/2\Gamma}\biggr)\biggr
  |_{0}^{\infty}\\
\end{eqnarray}
The divergence at the lower end needs to be cut off and this is done
by recognizing that the lowest value of $k$ is $\mathcal{O}(L^{-1})$
and to the leading order the integral is
\begin{eqnarray*}
  \int \frac{\mathrm{d}^2
\vec{k}}{(2\pi)^2}\frac{s}{(s+2\Gamma k^2)(2\Gamma k^2)} &=& \frac{1}{8\pi \Gamma}
\ln \biggr(1+{sL^2 \over 2\Gamma} \biggr) \\
   &=& \frac{1}{8\pi \Gamma} \ln \biggr({sL^2 \over 2\Gamma}
   \biggr) \phantom{1cm} (\mathrm{for}{sL^2 \over 2\Gamma}\gg 1)
\end{eqnarray*}
Consequently for $L$ finite but much greater than $s^{-1/2}$, the
expression for $\triangle \tilde{J}$ becomes
\begin{eqnarray}\label{20}
\nonumber
  \triangle \tilde{J}&=& {L\over 8\pi\Gamma}\biggr({s\over2\Gamma}\biggr)^{1/2}-
  \frac{1}{8\pi \Gamma} \ln \biggr({sL^2 \over 2\Gamma}
   \biggr)\\
\nonumber
   &=&{L\over 8\pi\Gamma}\biggr({s\over2\Gamma}\biggr)^{1/2} \biggr[
   1-{\ln \bigr({L^2s\over 2\Gamma}\bigr)
   \over L\bigr({s \over 2\Gamma}\bigr)^{1/2}}\biggr]\\
\end{eqnarray}
We now need to explore the limit $L({s\over 2\Gamma})^{1/2}\ll 1$.
To do this we return to Eq.(\ref{17}), perform the two dimensional
$k$ integration and write
\begin{eqnarray}\label{21}
\nonumber
\triangle \tilde{J}&=&{\pi\over2(2\pi)^2} \sum_n
\ln\biggr({n^2 \pi^2 \over L^2}+{s\over2\Gamma}\biggr)\biggr/{n^2\pi^2\over L^2} \\
\nonumber
   &=&{\pi\over2(2\pi)^2}\sum_n \ln \biggr( 1+{L^2 s\over 2n^2 \pi^2 \Gamma}\biggr) \\
   &=& {L^2 s\over 96 \pi \Gamma}
\end{eqnarray}
Since the denominator of Eq.(\ref{15}) already contains a term
linear in $s$ this limit $\triangle \tilde{J} $ will not reveal any
additional feature. For small value of $s$, we can now write for
$L({s\over 2\Gamma})^{1/2}\gg 1$
\begin{eqnarray}\label{22}
\nonumber
\tilde{g}(s)&=&{2\pi \over u\Gamma L(s/2\Gamma)^{1/2}}
\biggr[1+{\ln(L^2 s/2\Gamma)\over L(s/2\Gamma)^{1/2}}\biggr]  \\
   &=&{2\pi\over u\Gamma} \biggr[\biggr({2\Gamma \over L^2 s}\biggr)^{1/2}
   +\biggr({2\Gamma \over L^2 s}\biggr)
   \ln\biggr({L^2 s \over 2\Gamma} \biggr) \biggr]
\end{eqnarray}
The real time behavior is obtained by inverting the Laplace
Transform of $\tilde{g}(s)$ and we have
\begin{eqnarray}\label{23}
\nonumber
  g(t)&=&{2\pi\over u\Gamma} \biggr[{1\over\sqrt{\pi}} \biggr({2\Gamma \over L^2 t}\biggr)^{1/2}
  +{2\Gamma \over L^2} \ln\biggr({L^2\over 2\Gamma t}\biggr) \biggr]  \\
  \nonumber
   &=&{\sqrt{\pi}\over u\Gamma} \biggr({2\Gamma \over L^2
   t}\biggr)^{1/2} \biggr[ 1+\sqrt{\pi}\biggr({2\Gamma \over L^2
   t}\biggr)^{1/2} \ln\biggr({L^2\over 2\Gamma t}\biggr) \biggr] \\
   &=& {C \over t^{1/2}}\biggr[ 1+\sqrt{\pi}\biggr({2\Gamma \over L^2
   t}\biggr)^{1/2} \ln\biggr({L^2\over 2\Gamma t}\biggr) \biggr]
\end{eqnarray}
At this order the expression for $a(t)$ and $b(t)$ becomes
\begin{equation*}
 2b(t)={1\over 2}\ln t - \ln\biggr[1+\sqrt{\pi}
 \biggr({2\Gamma t \over L^2}\biggr)^{1/2}\ln\biggr({L^2 \over 2\Gamma t}\biggr)\biggr]
\end{equation*}
and
\begin{eqnarray}\label{24}
\nonumber
a(t)&=&{1 \over 4t} -{\sqrt{\pi}\over 4}{\biggr({2\Gamma
\over L^2 t}\biggr)^{1/2}\biggr(\ln\biggr({L^2 \over 2\Gamma
t}\biggr) -2\biggr)
 \over 1+\sqrt{\pi}\biggr({2\Gamma \over L^2
   t}\biggr)^{1/2} \ln\biggr({L^2\over 2\Gamma t}\biggr)}  \\
\nonumber
  &=&{1 \over 4t}-{\sqrt{\pi}\over 4}\biggr({2\Gamma
\over L^2 t}\biggr)^{1/2}{\ln\biggr({L^2 \over 2e^2\Gamma t}\biggr)
\over 1+\sqrt{\pi}
 \biggr({2\Gamma t \over L^2}\biggr)^{1/2}\ln\biggr({L^2 \over 2\Gamma
 t}\biggr)}\\
\end{eqnarray}
We note that for $L$ large enough so that ${L^2 \over \Gamma t} \gg
1$, $a(t)\simeq {1\over 4t}$, with the first correction given by
\begin{eqnarray}\label{25}
\nonumber
  a(t) &=& {1\over 4t}-\sqrt{\pi} \biggr({2\Gamma \over L^2 t}\biggr)
  \ln\biggr({L^2 \over 2e^2\Gamma t }\biggr) \\
  &=& {1\over 4t}\biggr[1-\sqrt{\pi} \sqrt{2\Gamma t \over L^2}
  \ln \biggr({L^2 \over 2e^2\Gamma t} \biggr) \biggr]
\end{eqnarray}
\\
For ${\Gamma t\over L^2}\ll 1$, we can write $a(t)$ as ${\epsilon(t)
\over 4t}$, where $\epsilon (t)$ is the quantity in brackets in
Eq.(\ref{25})and is slowly varying function in the range considered.

The global mode $\phi_1(0)$ now satisfies the equation of motion
(see Eq.(\ref{9}),
\begin{equation}\label{26}
\biggr({\mathrm{d}\over \mathrm{d}t } +{\Gamma \pi^2 \over L^2}
\biggr)\phi_1(0,t)={\epsilon(t) \over 4t} \phi_1(0,t)+\xi(t)
\end{equation}
Under the transformation $\phi_1(0,t)=e^{-\Gamma t\pi^2 /L^2}
t^{\epsilon(t)/4} \psi(t)$ and making the slowly time varying
approximation whereby $(\dot{\epsilon}/ \epsilon )t\ln t$ is
considered significantly smaller than unity (that is $\Gamma t/L^2$
reasonably smaller than unity) we arrive at
\begin{equation*}
\dot{\psi}(t)=e^{\Gamma t\pi^2 /L^2} t^{-\epsilon(t)/4} \xi(t)
\end{equation*}
With the transformation of variable $\tau=t^x$ we get
\begin{equation}\label{27}
\dot{\psi}(\tau){\mathrm{d}\tau \over \mathrm{d}t}=e^{\Gamma t\pi^2
/L^2} t^{-\epsilon(t)/4} \xi(t)=\tilde{f}(\tau)
\end{equation}
The correlation function $<\tilde{f}(\tau)\tilde{f}(\tau ')>$ will
be delta correlated in $\tau$-space provided
\begin{equation}\label{28}
x=1-{\epsilon \over 2}- {2\pi^2 \Gamma t \over L^2}{1\over
\ln(\Gamma t/L^2)}
\end{equation}
and Eq.(\ref{27}) becomes
\begin{equation}\label{29}
\dot{\psi}(\tau)=\tilde{f}(\tau)
\end{equation}
Since the size dependent correction in $\epsilon$ is
$\mathcal{O}(L^{-1})$, we can drop the last term to the leading
order and write as the first effect of the finite size, the relation
\begin{equation}\label{30}
 x={1\over 2} +{\sqrt{2\pi}\over 2}\sqrt{{\Gamma t\over L^2}}
 \ln \biggr({L^2 \over e^2 \Gamma t}\biggr)
\end{equation}
The persistence probability for the process of Eq.(\ref{29}) goes as
$\tau^{-1/2}$ and hence in the actual time variable $t$,
\begin{equation}\label{31}
p(t)\sim {1\over t^{{1\over 4}+\sqrt{{\pi\Gamma t \over 8
L^2}}\ln(L^2/e^2\Gamma t)+...}}
\end{equation}
The decay is clearly hastened at a finite value of L.

What happens is $L^2/\Gamma t$ becomes smaller than unity? Returning
to Eq.(\ref{15}) and Eq.(\ref{21}), it is now clear that the leading
behavior of $g(t)$ is $e^{-\Gamma \pi^2 t/L^2}$ leading to
$b(t)={\Gamma \pi^2\over L^2}t$ and $a(t)={\Gamma \pi^2 \over L^2}$.
This implies a dynamics
\begin{equation}\label{32}
{\mathrm{d} \over \mathrm{d}t} \phi_1(0,t)=-{\Gamma \pi^2 \over L^2}
\phi_1(0,t) +\xi(t)
\end{equation}
The associated $p(t)$ is known from ref.\cite{8} to be
\begin{equation}\label{33}
p(t)=\sqrt{{\Gamma \pi^2 \over L^2}} {e^{{-\Gamma \pi^2 t\over L^2
}}\over \sqrt{\sinh({\Gamma \pi^2 t \over L^2})}}
\end{equation}
A combination of the forms of Eq.(\ref{31}) and Eq.(\ref{33}) can be
achieved by
\begin{equation}\label{34}
p(t)={e^{{-\Gamma \pi^2 t\over L^2}} \over \biggr \{{L^2 \over
\Gamma \pi^2}\sinh({\Gamma \pi^2 t \over L^2})\biggr\}^{{1\over4}
+{1\over 4+\alpha}}}
\end{equation}
where
\begin{equation}\label{35}
\alpha=\sqrt{{8L^2\over\pi\Gamma t}}{1\over\ln({L^2 \over e^2\Gamma
t})}
\end{equation}
For $L^2\gg\Gamma t$, we have the result of Majumdar et. al.
\cite{1}, that is $p(t) \sim t^{-1/4}$, while for $L^2 \ll \Gamma
t$, we regain Eq.(\ref{33}).

We note that the critical relaxation rate for a system governed by Eq.(\ref{2}) goes as $\Gamma k^2$, where $k$ is the wavenumber of the fluctuations. For the finite sized system the minimum value of $k$ is $\mathcal{O}(L^{-1})$ and the lowest frequency is given by $\Gamma L^{-2}$ . For any arbitrary time scale $'t'$, the relaxation rate allows us to define a dynamic length scale $l_d =\sqrt{\Gamma t}$. The ratio $L^2/\Gamma t$ which has featured so prominently in our discussion is thus the ratio $L^2/l^2_{d}$. The results of Majumdar et al \cite{1} are for the limit $L\gg l_d$. Our Eq.(\ref{34}) is an attempt to capture the entire range from $L\gg l_d$ and $L\ll l_d$.

{\bf{\large{Acknowledgment:}}}\\
D.C acknowledges Council for Scientific and Industrial Research, 
Govt. of India for financial support (Grant No.- 9/80(479)/2005-EMR-I).

\end{document}